\documentclass[a4paper,12pt]{article}
\usepackage{amssymb}
\usepackage{epsfig}
\usepackage{amsmath}
\usepackage[english]{babel}

\author {Yuriy\,A.\,Bunyak\/\thanks{bunyak@innovinn.com}
\\{ \small IVP InnoVinn, Kyjivska,14, 21100 Vinnitsa, Ukraine}
}
\title{Numerical analysis of solitary waves interaction in nonlinear medium} 

\date{\nonumber}
\begin{document}

\newcommand{\sech}{\rm sech}

\maketitle

\begin{abstract}
{Using numerical modeling investigated interaction of solitary waves (solitons) 
of the regularized long wave equation. For reception the stable model of the 
nonlinear medium are used methods of the linear prediction and progressive 
approximation. By modeling was determined that depending on ratio of velocities 
of the solitons and the form of highest derivatives balance is possible 
self-organization of the medium nonequilibrium state as formation of shock 
waves and stable on the form solitary waves, created as a result of full or 
partial mutual penetration of the solitons. Is possible also aggregation of 
the solitons in third wave. The shock waves can pass into other possible 
resonance state as a wave front with stable amplitude, which precedes developing 
in singularity negative front.}
\end{abstract}

Spatially separated solitary waves (SW) are in balance with nonlinear medium 
if they are resonance to it. At rapprochement of the SW the balance is saved when 
multiwave field is resonance too. Otherwise the wave field dynamic is difficult 
predictable. An interaction of the solitary waves of the nonresonance wave field 
is possible to analyse by numerical modeling. As analyse object we shall choose 
nonlinear medium, described by the regularized long wave (RLW) equation 
\begin{align}\label{eq:1}
u_t+\gamma u_x-\alpha u_{xxt}+2\beta uu_x=0,
\end{align}
where $u=u(x,t)$, $\gamma =1$, subscript symbols mark the partial derivatives on spatial 
$x$ and temporal $t$ variables. The RLW equation describes haracteristic of the nonlinear 
medium in more broad dynamic range in contrast with Korteveg - de Vries's (KdV) equation, 
however, it is't solvable by the inverse scattering method and other algebraic methods and 
so has't ~{$N$-solitons} solutions. The RLW equation is considered in literature with pairs 
of coefficients: $(\alpha;\beta)=(1;0.5)$ ~\cite{bib:LA}, (1;-3) ~\cite{bib:LB} and 
(0.5;0.125) ~\cite{bib:LC}. Substitution ~$u=w_x$ and Hopf - Cole's 
transformation ~\cite{bib:LA} as ~{$w=\lambda(\ln\Gamma)_x$}, where 
$\Gamma(x,t)=exp(a(x-vt))+1$, $\lambda$, $a$ -- constants, $v$ - velocity, yield
the condition of the highest derivatives balance ~$6\alpha\Gamma_t+\lambda\beta\Gamma_x=0$, 
from which follows that the velocity 
\begin{equation}\label{eq:3}
v=\lambda\beta/6\alpha.
\end{equation} 
From linear part ~(\ref{eq:1}) follows that spectral parameter 
$a=\sqrt{(v-\gamma)/\alpha v}$. The solution of the RLW equation with account of the 
expression for velocity is
\begin{equation}\label{eq:2}
u(x,t)=\frac 3{2\beta}(v-\gamma)\sech^2\left(\frac 12\it \sqrt{\frac{v-\gamma}{\alpha v}}(x-vt)\right).
\end{equation}
When ~{$v>\gamma$} the function $u(x,t)$ describes the solitary wave, which on nature 
of the dependencies of the amplitude from velocity, locality and conservation of the 
form at spatio-temporal translations corresponds to the determination of the soliton 
~\cite{bib:LA}. When ~$v<\gamma$ the solution 
\begin{equation}\label{eq:4}
u(x,t)=\frac 3{2\beta}(v-\gamma)\cos^{-2}\left(\frac 12\sqrt{\frac{\gamma-v}{\alpha v}}(x-vt)\right) 
\end{equation}
is periodically singular. We shall research an interaction of the one-soliton 
solutions ~(\ref{eq:2}) for different ratios of the velocities and pairs of the 
coefficients $(\alpha,\beta)$ using numerical simulation of the dynamic of an 
initial layer, defined as two waves, given on a square grid with intervals of 
sampling ~$\Delta x$,~$\Delta t$ on space and time accordingly.   

The numerical modeling of nonlinear processes is characterized ambiguous results, 
since presence power-mode function implies minimum two decisions. If chosen 
decision will not matched on the dynamic with previous, that the numerical scheme 
can lose stability. Since a nonlinearity reveals itself as influence of a sought 
value on itself, that appears the problem of the choice of the supporting area 
for generation of following values of the net under influence not yet defined 
value. Therefore computing model of the nonlinear process must be built on scheme of 
the progressive approximation with restriction on area of possible result values.
 
Invariance of the solitons to spatio-temporal translation is expressed by equation
\begin{equation}
u_t+vu_x=0\label{eq:5}
\end{equation}
with the dispersion equation ~{$\omega-vk=0$}. Let $u_{i,j}=~{u(j\Delta x,i\Delta t)}$ 
denotes the samples of the discrete wave field, where $i,j=0,1,\dots\,$, and 
~{$v_0=\Delta x/\Delta t$} is the velocity of the moving wave, matched with the samples 
net. As follows from equation ~(\ref{eq:5}), for the soliton with velocity ~$v=mv_0$, 
where $m$ -- integer number (the velocity index), is obviously identity 
~$u_{i,j}\equiv u_{i-1,j-m}$. By applying ~{$z$-images} of the shift on time $z_t$ and 
space $z_x$ instead of finite differences the discrete analogue of the equation ~(\ref{eq:5}) 
can be written as $1-z_t{z_x}^m=0$. For two solitons, moving with multiple $v_0$ 
velocity ~{$v_1=m_1v_0$} and ~{$v_2=m_2v_0$}, the ~{$z$-image} of the dispersion 
equation ~{$(\omega-v_1k)(\omega-v_2k)=0$} transforms to 
$1-z_tz_x^{m_1}-z_tz_x^{m_2}+ z_t^2z_x^{m_1+m_2}=0$ and the corresponding difference 
scheme $u_{i,j}-u_{i-1,j-m_1}-u_{i-1,j-m_2}+u_{i-2,j-m_1-m_2}\equiv 0$. For wave that 
containing $N$ solitons from ~{$z$-presentations} of the dispersion equation is possible 
to form following difference scheme.
\begin{eqnarray}
\sum_{n=0}^N\sum_{l=0}^M\,a^0_{n,l} u_{i-n,j-l}\equiv0;\label{eq:6}\\
a^0_{0,0}=1;\,a^0_{0,1\dots M}=0;\,a^0_{n,p}=(-1)^n\delta_{p,S};\nonumber\\
S=\sum_{l=1}^n m_{\sigma_l};\,\sigma\subset m_1\dots m_N;\nonumber\\
n=1,2,\dots,N;\,p=1,2,\dots,M,\nonumber
\end{eqnarray}
where $M$ is the sum of $N$ velocity indexes ~$m_{\sigma_l}$. If $p$ in
~(\ref{eq:6}) is equal to the sum of $n$ velocity indexes from given ensemble 
$N$ indexes, then ~{$\delta$ - function} is an unit, otherwise it's a zero.

The double sum in expression (\ref{eq:6}) presents itself equation of the 
two-dimention linear prediction (LP) ~\cite{bib:LD} -- samples of ~$u_{i,j}$ 
are possible to define as linear combination of previous, given as matrix of size 
~$(N+1)\times(M+1)$. For $N$ solitons with arbitrary velocity indexes the  
LP model is given by
\begin{equation}
\sum_{n=0}^N\sum_{l=0}^M\,a_{n,l} u_{i-n,j-l}\approx 0;\,a_{0,0}=1,\label{eq:7}
\end{equation}
where $M$ is rounded value of the sum of $N$ velocity indexes. The LP parameters  
~$a_{n,l}$ may be defined using the least square method on data of the field 
in initial layer. 

The derivatives on $x$ and $t$ can be  presented as finite differences with use 
the equation ~(\ref{eq:7}). 
\begin{alignat}{2}
\frac1{\Delta t}(u_{i,j}-u_{i-1,j})=\frac1{\Delta t}\left(u_{i,j}+\frac{u_{i,j}}{a_{1,0}}+\frac{S_{1,0}}{a_{1,0}}\right),\label{eq:8}\\
\frac1{\Delta x}(u_{i,j}-u_{i,j-1})=\frac1{\Delta x}\left(u_{i,j}+\frac{u_{i,j}}{a_{0,1}}+\frac{S_{0,1}}{a_{0,1}}\right),\label{eq:9}
\end{alignat}
where
\begin{displaymath}\nonumber
S_{p,q}={\sum_{n=0}^N\sum_{l=0}^M}_{(n,l)\neq(0,0),(p,q)}\,a_{n,l} u_{i-n,j-l},
\end{displaymath}
indexes in brackets denote the pair values. Such approach allows to increase
the supporting area for calculation of the finite differences, as well as allows 
to normalize supporting areas of differences first and following orders.
Using similar ~(\ref{eq:8}), ~(\ref{eq:9}) presentations of the high order derivatives 
for the equation ~(\ref{eq:1}) shall receive following iteration difference scheme.
\begin{equation}\label{eq:10}
u_{i,j}(k)=-\left(C_0{u_{i,j}}^2(k-1)+C_2\right)/C_1,
\end{equation}
where $k$=0,1,\dots\,-- iteration step number;
\begin{displaymath}\nonumber
\begin{split}
u_{i,j}(0)=-S_{0,0};\\
C_0=\beta\frac{\Delta t}{\Delta x}\left(1-\frac1{a_{0,1}^2}\right);\\
C_1=1+\frac1{a_{1,0}}-
\frac{\alpha}{\Delta x^2}\left(1+\frac2{a_{0,1}}-\frac1{a_{0,2}}+\frac1{a_{1,0}}-\frac2{a_{1,1}}+\frac1{a_{1,2}}\right)+
\frac{\Delta t}{\Delta x}\left(\gamma+\frac{\gamma}{a_{0,1}}-2\beta\frac{S_{0,1}}{a_{0,1}^2}\right);\\
C_2=\frac{\alpha}{{\Delta x}^2}\left(-\frac{2S_{0,1}}{a_{0,1}}+\frac{S_{0,2}}{a_{0,2}}-\frac{S_{1,0}}{a_{1,0}}+\frac{2S_{1,1}}{a_{1,1}}-\frac{S_{1,2}}{a_{1,2}}\right)+
\frac{S_{1,0}}{a_{1,0}}+\frac{\Delta t}{\Delta x}\left(\gamma\frac{S_{0,1}}{a_{0,1}}-\beta\frac{S_{0,1}^2}{a_{0,1}^2}\right).
\end{split}
\end{displaymath}
In equation ~(\ref{eq:10}) the nonlinear component is presented as discrete 
analogue of $\beta (u^2)_x$. As can be seen from ~(\ref{eq:10}), the LP coefficients 
must be nonzero. For this the velocity indexes of the waves follow to choose 
as rational numbers with fractional part not multiple $2^{-n}$. 

The numerical dynamic model of the initial layer allows to consider the 
interaction of solitons, moving in forward direction. The study of the models 
~(\ref{eq:7}), ~(\ref{eq:10}) has shown that the model ~(\ref{eq:7}) reproduces 
the one-soliton wave with error linearly increasing from layer to layer, 
which relative value on a border of the net of size $500\times 500$ not 
more $0.001$. The LP model reproduces two and more solitons with relative 
inaccuracy not more $0.03$, moreover, the inaccuracy value sharply increases 
in cross point of the solitons. Since given inaccuracy is a smooth 
function, reveals itself on solitons fronts and does't influence upon their 
form, that it possible consider as effect of the phase shift. The model 
~(\ref{eq:10}) reproduces the initial layer of the one-soliton wave with small 
phase shift. The iterative procedure was checked by convergence condition 
$\mid u_{i,j}(k)-u_{i,j}(k-1)\mid \leq 1.e-10$ and was limited by ten steps. 
Performing the convergence condition was fixed in all nodes of the net. 
The studies have shown that important is the choice of the model order in 
accordance with ~(\ref{eq:6}) -- ~{$z$-image} of ~(\ref{eq:7}) must corresponds 
to the product of $N$ dispersion equations ~(\ref{eq:5}) each of solitons.
Stability given model depends on magnitude of the LP coefficients, that depend 
on the choice of the sampling interval and velocity scaling. Is possible also 
the correction of the LP model order for reason to remove critical magnitude 
of the coefficients.

On ~{Fig.1} is presented the interaction of two solitons of the RLW equation 
with parameters (0.5;0.125), when influence of nonlinearity is least. 
The velocities ratio less 1.2\,. The model parameters are following: ~{$v_0=10$}; 
~{$\Delta x=0.25$}; ~{$N=3$}; 
~{$M=2$}; ~{$m_1=0.605$};~{$m_2=0.67$}. Before cross point solitons save the form, 
their amplitudes summarize linearly. Further occurs qualitative change of the 
wave form -- are formed quick waves with slowly rising amplitude and a shock wave.
Simultaneously with growing of the amplitude and narrowing the profile of the shock 
wave is formed a negative front, an amplitude which sharply increases after  
the positive front achieved a fixed level. The negative front moves with greater 
velocity and with increase the distance from saving form positive front it gradually 
forms a surge in area of great numbers. On correlation of the amplitude and velocity 
of the positive front, development of the negative front dynamic, smoothly moving 
over to singularity, possible draw the conclusion that the shock wave is transformed 
in a wave like ~(\ref{eq:4}) accurate to phase shift and scaling multiplier on 
velocity and amplitude. The numerical decision converges in accordance with presented 
above criterion in all nodes of the net. The integral value under profile of 
the wave was saved. Only in the field of sharply growing of the negative front 
the iterative process falls into closed cycle around neighbour values. If consider
~(\ref{eq:10}) as quadratic equation then in this area its discriminant negative. 
Therefore the model in this area has evaluating nature. The form of singular area 
greatly is't changed with increase of quantity of iteration steps. 

Within the range of the velocities ratio 1.2\dots 1.25 the solitons interaction carries 
the elastic nature -- the solitons disperse with conservation of the form and velocity. 
Given interaction possible to characterize and as a full mutual penetration, since 
in the cross point amplitudes of solitons summarize linearly. Similarly interact solitons 
with velocity indexes ~{$m_2=(K\pm 0.25)m_1$}, where $K$ is integer value.

On ~{Fig.2} is presented an instance of the solitons interaction with the velocities 
ratio outside of noted above areas: ~{$m_1=0.55$};~{$m_2=0.7$}. The interaction 
occurs inelastically. On ~{Fig.3} is presented the consequence of shears of the wave 
before and after the interaction. From figures is seen that approximately only about 
half on amplitude and length of the quick soliton continues the motion with the previous 
velocity as the solitary wave. The length of this wave is a half of the length of 
the slow soliton. The slow soliton absorbed part of the quick soliton energy 
and transformed in the shock wave with the same scenario of the behaviour, as on ~{Fig.1}. 

If consider solitons as a sum of harmonical waves then from presented results 
follows that exist tunnel zones in neighborhood of multiple magnitudes of the velocities, 
when harmonic components mutually penetrate. Outside of the tunnel zones more slow 
wave emerges as potential barrier, through which passes the part of harmonical 
waves, period which is multiple with respect to length of the barrier. Rest harmonical 
waves enter in nonlinear interaction with the barrier and create the shock wave. Part 
of waves, which passed the barrier, creates stable on the form solitary wave, moving 
with previous velocity. As can be seen from the expression ~(\ref{eq:3}), such wave 
can be considered as solution of the equation ~(\ref{eq:1}) with proportionally increased 
coefficients $(\alpha;\beta)$. Consequently, resonance to the nonlinear medium can 
be SW, which parameters give appropriate balance of the form and velocity.

In the equation ~(\ref{eq:1}) with parameters ~{(1;0.5)}, ~{(1;-3)} the nonlinear 
component has more deep influence. Therefore mutual penetration of the waves with small 
distortion was observed only in the tunnel zones with $K \geq 2$. On ~{Fig.4} is 
presented the solitons interaction of the RLW equation (1;0.5) with parameters of 
the model: ~{$N=2$}; ~{$M=3$}; ~{$m_1=0.6$}; ~{$m_2=0.7$}. The interaction runs 
with formation before front of the slow soliton the third wave, which gradually 
accumulates the main part of total energy of the solitons and develops in the shock 
wave, and next in wave like ~(\ref{eq:4}). In this case the transition in 
the wave like ~(\ref{eq:4}) is realized quicker and so the shock wave excites 
the series of the fluctuations developing in periodical singularites. 
The third wave is formed as reflection of the slow wave from running up front of 
the quick wave. Then reflected wave absorbs quick soliton. Such interaction shows 
that with increase the influence of a nonlinearity grows the ability of the solitons 
to save the form. The interaction of the solitons of the RLW equation with 
parameters (1;-3) occurs similarly and differs the inverse sign of the wave.

Presented in article the model has allowed to see result of influence of the 
nonlinearity on non-equilibrium interaction of the SW. As a groundwork to influence 
is used the linear model (\ref{eq:7}), which is equivalent to superposition of the 
operators (\ref{eq:5}) each of SW. 
\begin{displaymath}\nonumber
\left\{\prod_{n=1}^N\left(\frac \partial{\partial t}+v_n \frac \partial{\partial x}\right)^{p_n}\right\} u(x,t)=0,
\end{displaymath}
where $p_n$ are integer values, which regulate the filtering characteristic 
each of multiplier. For realization of the computing scheme ~(\ref{eq:10}) the 
auxiliary linear model can be formed for each layer or counting sample on its 
neighbourhood. The linear model must be unitary to not influence on processes 
of damping and excitation of fluctuations.
With the help of linear model was eliminated ambiguity of nonlinear 
equation solution by matching on a dynamic of current and previous samples. 
Model was researched also with discrete analogue of the nonlinearity as 
$u_{i,j-1}(u_{i,j}-u_{i,j-2})/2\Delta x$. Such model gave the results, similar
mentioned above, but it was less stable in contrast with ~(\ref{eq:10}) and 
amplitude of the shock wave did't rise up to saturation level.

Possible suppose that transformation of the shock wave in the wave like 
~(\ref{eq:4}) occurs under influence of $\gamma u_x$ in equation ~(\ref{eq:1}). 
To value the level of this factor the model, presented on ~{Fig.1}, was researched 
for $\gamma=-0.25 \dots 1.5$. It was determined that ratio of the amplitude
of the shock wave saturation to the average magnitude of the solitons amplitudes  
is valued as about $24-8\gamma$ and though amplitudes of the solitons 
with reduction $\gamma$ increase the time of the shock wave saturation grows 
as about $T(\gamma=1)+160\Delta t(1-\gamma)$. These results show that $\gamma$ 
affects only partly -- with the reduction of $\gamma$ decreases ability to 
saturation. Since a shock wave with high amplitude is't a solution of the RLW 
equation with any parameter $\gamma$, that the main factor is a process of 
self-organization of the shock wave under influence of the factors of amplitude, 
frontage and velocity, in consequence of which the shock wave shapes up the 
resonance form  that is acceptable from condition of the high derivatives balance. 
This is confirmed following, when the solitons amplitude was changed by parameter 
$v_0=10\dots 200$ the ratio of the amplitude of the shock wave saturation  
to average magnitude of the solitons amplitudes was valued as about 
$8{v_0}^{0.3}$ and the time of saturation greatly was't changed. However,
was found ranges where does't occur the shock wave saturation: 
$v_0=70\dots 80$; 125\dots 150; 180\dots 200. The shock waves 
with high amplitude are formed in these ranges right after collision, 
or in cross point of the SW. Whereupon the model lost stability. In addition, 
the double order was needed for stable linear model in these ranges. The 
model of the KdV equation behaved similarly.

When $v\gg\gamma$ the spectral parameter in equation ~(\ref{eq:2}) 
$a\approx\sqrt{1/\alpha}$ and so solitons of the RLW equation are spatially 
quasicoherent. 
Hence, the nonequilibrium interaction of the quasicoherent solitons can generates 
solitary and shock waves, self-excited quick waves-harbingers. The shock wave can 
reach to be tenfold amplitude in contrast with generating it solitons due 
to the profile narrowing. In the case of weak nonlinearity possible 
partial mutual penetration with formation of saving form solitary waves. 
Strong nonlinearity brings about forming the third wave, which accumulates 
the main part forming it the solitons. The self-excitation of the quick 
waves-harbingers can be considered as property of a nonlinear medium 
spontaneously to generate fluctuations under influence of small perturbations.

The model has reflected conversion of the nonlinear medium from one type of the 
balance of the amplitude and velocity -- two solitons ~(\ref{eq:2}), to the 
other, in the form of the stable positive wave front, which precedes developing in   
singularity negative front. Physical sense of such transition possible to make clear 
considering other characteristic of concrete phenomena, described by 
equation ~(\ref{eq:1}). For example, the RLW equation (0.5;0.125) describes an 
ion velocity in Broer - Sluijter's system of equations of plasma ~\cite{bib:LE}. 
Numerical modeling of the dynamic of the ion density at the speed on ~{Fig.1} using 
the direct matrix method has shown that when ion velocity takes the form of 
the shock wave with stable amplitude and appears negative front the amplitude 
of fluctuations of the ion density sharply increases.
The negative velocity possible to consider as effect of the elastic force reaction 
on quick ions. The rapprochement with the singular point of the velocity
brings about forming the fluctuations of the density with high amplitude, spreading 
in forward and backward directions. Consequently, the singularity corresponds to 
excited by colliding ions source of the fluctuations of density.  

Solitary waves can meet in nature as multiwave systems and as separate 
formations, depending on this condition runs their interaction. The first type of 
the interaction well studied by analytical way. Example of the numerical analysis 
of the second type of the interaction is presented in this article. Results of 
modeling allow to draw a conclusion that nonlinearity reveals itself ambiguously -- 
in one event exists the mutual penetration, in the other -- full or partial 
interaction with transition in qualitative other resonance state. Partly the 
interaction of the SW in nonlinear medium possible to describe qualitatively as 
processes of the multiplying and mutual passing -- reflections of their 
harmonic components taking into account orthogonality of the harmonics and 
conservations of total energy.

\onecolumn

\begin{figure}
\begin{center}
\includegraphics[height=8cm,width=16.0cm]{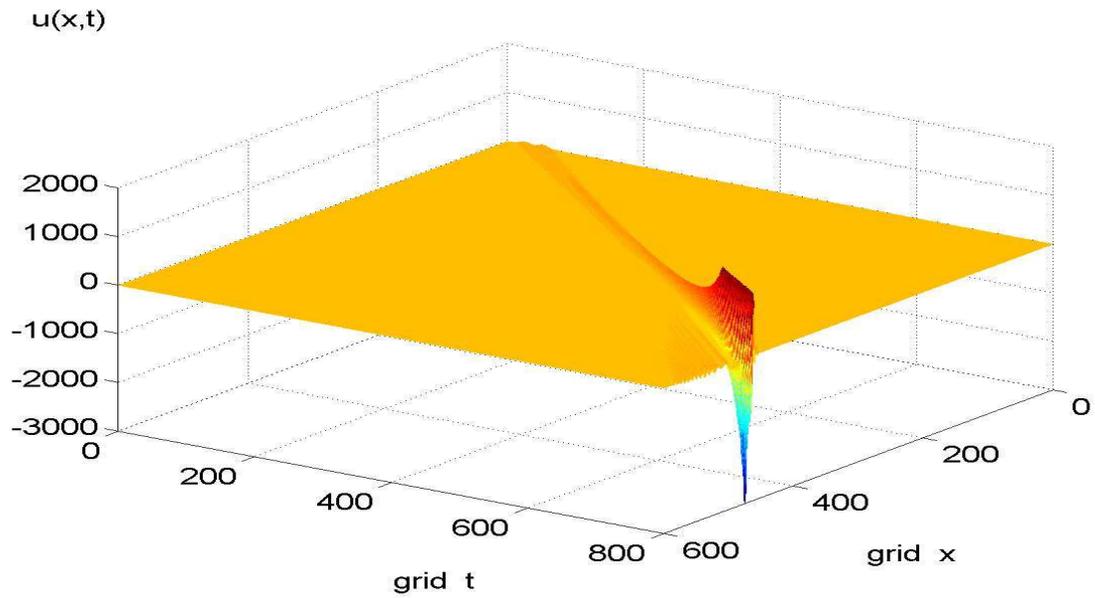}
\caption{Solitons interaction with parameters of the model: ~$\alpha=0.5;$ ~$\beta=0.125;$ ~$v_1=6.05;$ ~$v_2=6.7;$ ~$N=3;$ ~$M=2$.}
\end{center}
\end{figure}

\begin{figure}
\begin{center}
\includegraphics[height=8cm,width=16.0cm]{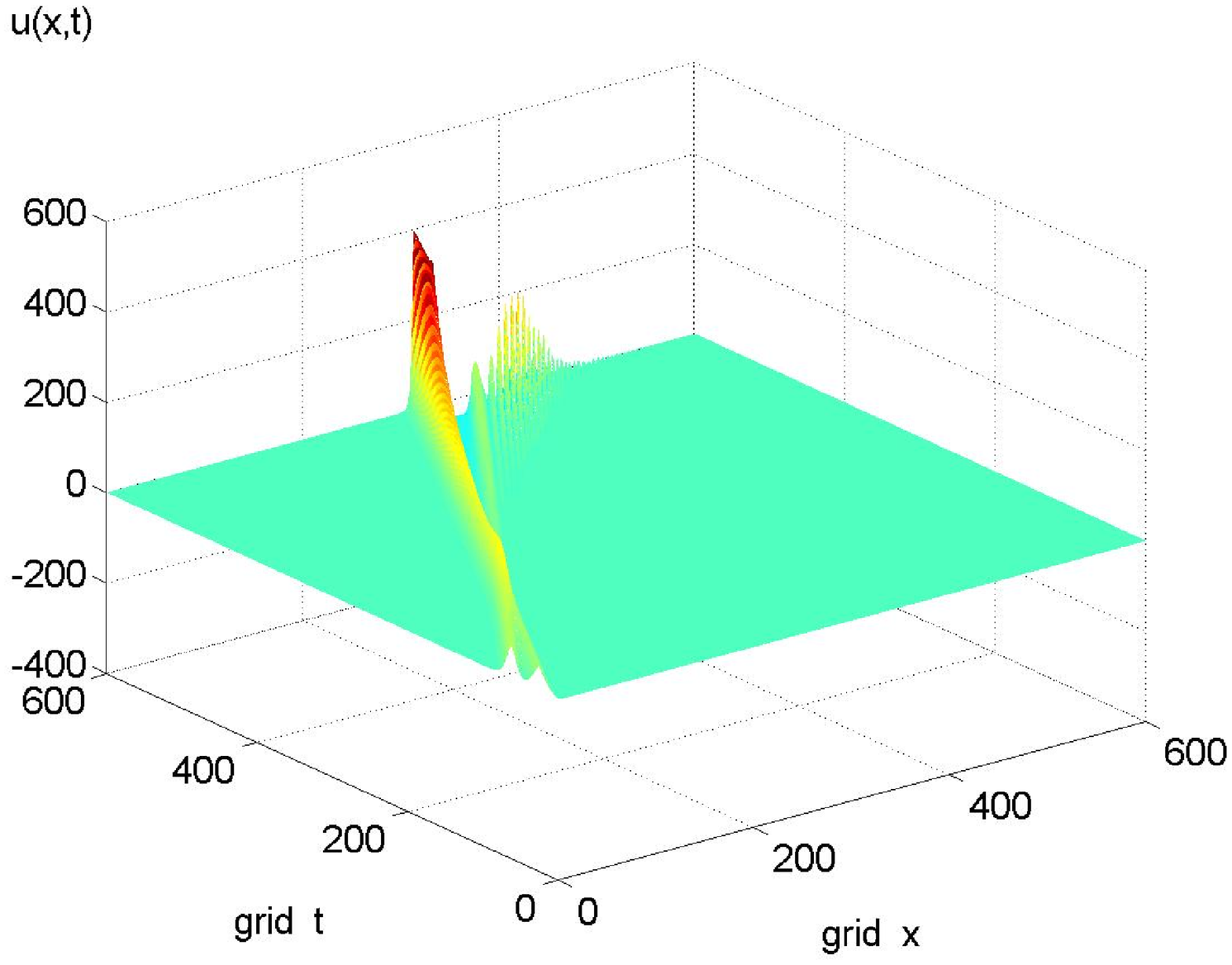}
\caption{Solitons interaction with parameters of the model: ~$\alpha=0.5;$ ~$\beta=0.125;$ ~$v_1=5.5;$ ~$v_2=7;$ ~$N=3;$ ~$M=2$.}
\end{center}
\end{figure}

\begin{figure}
\begin{center}
\includegraphics[height=8cm,width=16.0cm]{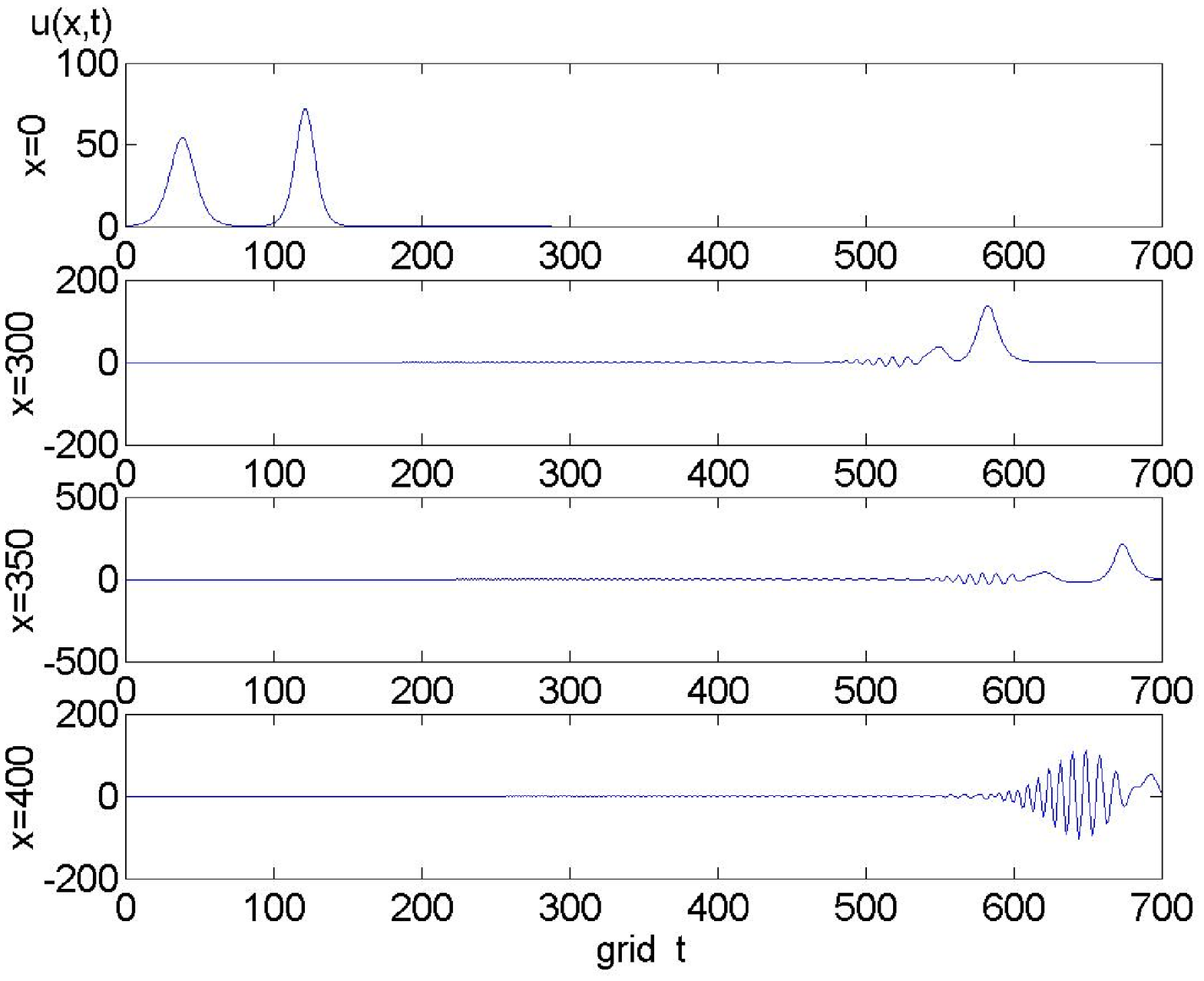}
\caption{Spatial shears of the solitons interaction with parameters of the model: ~$\alpha=0.5;$ ~$\beta=0.125;$ ~$v_1=5.5;$ ~$v_2=7;$ ~$N=3;$ ~$M=2$.}
\end{center}
\end{figure}

\begin{figure}
\begin{center}
\includegraphics[height=8cm,width=16.0cm]{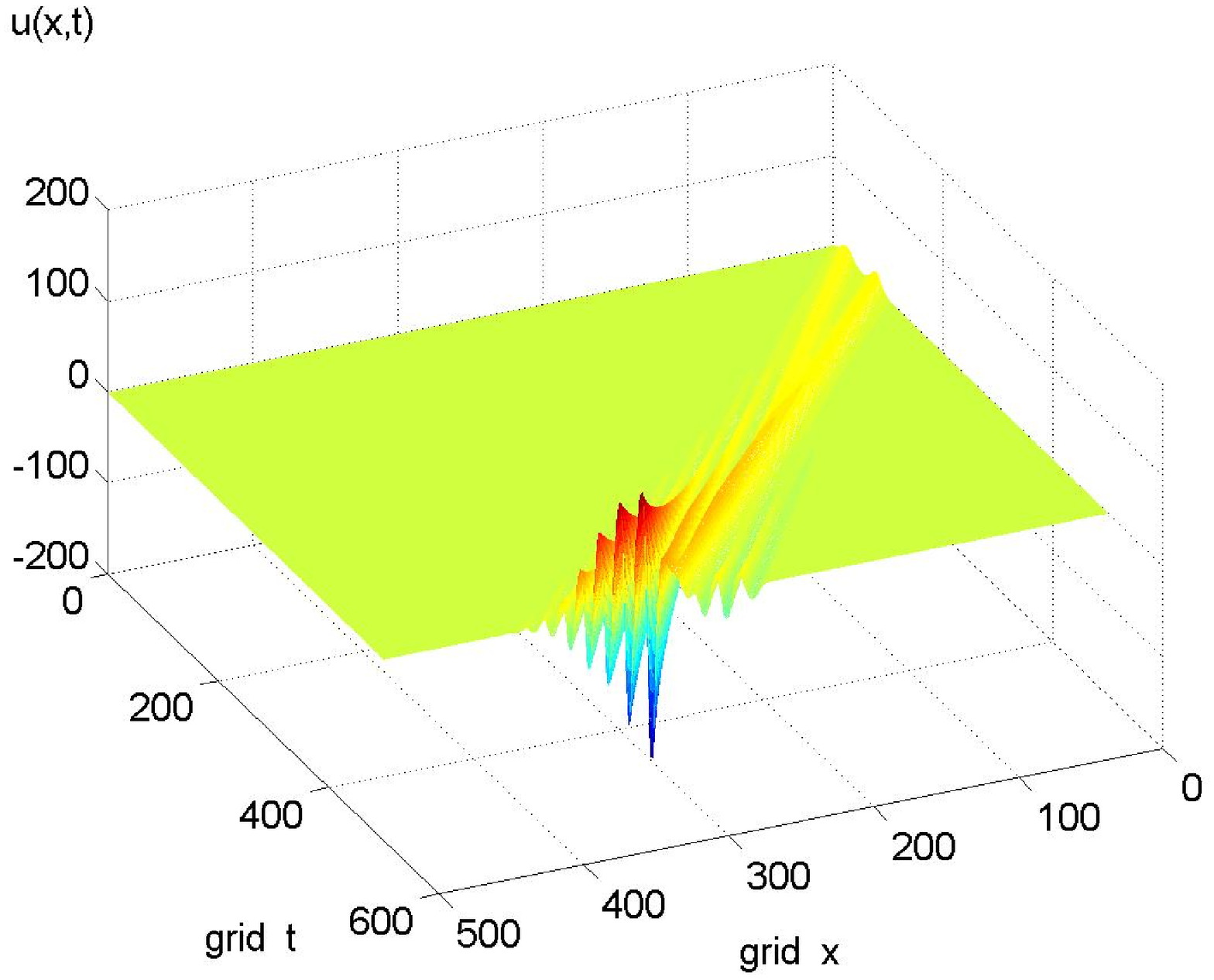}
\caption{Solitons interaction with parameters of the model: ~$\alpha=1;$ ~$\beta=0.5;$ ~$v_1=6;$ ~$v_2=7;$ ~$N=2;$ ~$M=3$.}
\end{center}
\end{figure}

\vfill\eject

\end{document}